\documentclass[reprint,%
aps,pra,%
10pt,%
superscriptaddress,%
amsmath,amssymb,%
floatfix]{revtex4-2}

\usepackage{amsmath,amssymb,bm}
\usepackage{mathtools}
\usepackage{esvect}
\usepackage{booktabs}
\usepackage{multirow}
\usepackage{array}

\usepackage{amsmath,amssymb,bm}
\usepackage{mathtools}
\usepackage{esvect}
\usepackage{booktabs}
\usepackage{multirow}
\usepackage{array}
\usepackage[svgnames]{xcolor}
\usepackage[breaklinks=true]{hyperref}
\begin{document}

\title{Decoherence of Atomic Ensembles in Optical Lattice Clocks by Gravity}
\author{Akio Kawasaki}
\email{akio.kawasaki@aist.go.jp}
\affiliation{National Metrology Institute of Japan (NMIJ), National Institute of Advanced Industrial Science and Technology (AIST), 1-1-1 Umezono, Tsukuba, Ibaraki 305-8563, Japan}

\begin{abstract}
Optical lattice clocks can now resolve the height difference below 1 mm within an atomic ensemble by means of gravitational redshift with integration over sufficient amount of time. Further improvement in the stability enables the clocks to resolve the height difference of subsystems within an atomic ensemble that is conventionally interrogated as a single coherent spin state in a single Ramsey sequence, resulting in the dephasing of the coherent spin state. This effect is observable with a clock of a stability of $\lesssim 10^{-21}$ by introducing a single-layer-resolved imaging system for a three-dimensional optical lattice, and limits the 1 s stability of the clock around $10^{-19}$ for an atomic ensemble distributing symmetrically for the three axes, which is also a signal of the decoherence. With atoms in an entangled state, this can be the first observation of the decoherence of a quantum state by a gravitational effect, and the suppression of other systematic shifts to observe this decoherence seems feasible. 
\end{abstract}

\maketitle

\section{Introduction}

Coherence is an important property of quantum states that enables interference of states specific to quantum mechanics. Although maintaining coherence for a long time is desired to make use of quantum states, decoherence is induced by various interactions with environments, such as fluctuating external fields, effects of impurities, collisions to background gases, and interactions between other quantum states nearby, depending on systems. One of the potentially ultimate sources of decoherence is the quantum mechanical fluctuation of the gravitational field, which is called gravitational decoherence \cite{JPhysConfSer.306.012006,GenRelGrav.28.581,PhysLettA.120.377}. Such decoherence of quantum states by gravitational effects has been analyzed as one of the paths towards unifying quantum mechanics and gravity. The basic analysis deals with a single qubit and analyzes the decoherence of a superposed state of this qubit due to the fluctuation of the metric \cite{ClassQuantumGrav.34.193002,PhysRevD.68.085006}. Some other reports analyze the decoherence bewteen multiple qubits due to gravitational interaction between them \cite{PhysRevA.95.052116,PNAS.114.E2303}. Decoherence of a single qubit under external field is also analyzed \cite{PhysRevA.95.052116}, but multiple qubits under external gravitational field has not been carefully dealt with, though recently developed optical lattice clocks provides an ideal experimental realization of such a system. 

Atomic clocks witnessed a significant improvement in their stability in past decades \cite{RevModPhys.87.637,LaRivNuovoCimento.12.555}. Multiple groups demonstrate the relative stability in the order of $10^{-18}$ with different atomic species\cite{NatPhoton.9.185,NatCommun.6.6896,Science.341.1215,PhysRevLett.116.063001,Metrologia.56.065004}, leading to the relative accuracy of $10^{-18}$. Comparisons of the stability between multiple atomic systems with a common laser systems reach the level of $10^{-19}$ \cite{Nature.588.408,NatPhoton.13.714,Nature.564.87}, and clocks with the $10^{-19}$ accuracy are appearing \cite{PhysRevLett.123.033201}. Among them, optical lattice clocks \cite{NatPhoton.9.185,NatCommun.6.6896,Science.341.1215,Nature.588.408,NatPhoton.13.714,Nature.564.87,Metrologia.56.065004} have an advantage of interrogating many, e.g. thousands of, atoms simultaneously, resulting in high stability in a fixed amount of time. Recent reports of strontium optical lattice clocks reach the stability below $10^{-19}$ \cite{2109.12237} and $10^{-20}$ \cite{2109.12238}. 

The long-lasting researches triggered several variations of optical lattice clocks. The most conventional one traps single atomic species in a one-dimensional optical lattice, and the atomic state is read out by collecting scattered light from the atomic ensemble with a single-channel photodetector such as a photomultiplier tube \cite{NatPhoton.9.185,Science.341.1215}. Currently, some clocks traps atoms in a two \cite{Nature.588.414} or three-dimensional \cite{Science.358.6359} optical lattice, or an array of optical tweezers \cite{PhysRevX.9.041052,Nature.588.408}. High resolution imaging systems \cite{PhysRevLett.120.103201,2109.12238} and single-atom-resolvable imaging systems \cite{PhysRevX.9.041052,Nature.588.408} brought a better understanding of the internal structure of atomic ensembles. Some optical lattice clock systems can trap multiple atomic species \cite{IEEETUFFC.65.1069} or multiple atomic ensembles \cite{2109.12237} in the same vacuum chamber, and the control and understanding of the environment, such as temperature \cite{PhysRevLett.113.260801}, electric field \cite{PhysRevLett.120.183201}, and magnetic field, have improved significantly. Also, combining an optical cavity and an optical lattice clock opened a way for further improvement in the stability of the state-of-the-art optical lattice clock with a help of quantum metrology \cite{Nature.588.414}. 

Various phenomena became detectable with these high precision clocks. One of them is systematic shifts and uncertainties \cite{PhysRevLett.113.260801,PhysRevLett.120.183201,PhysRevLett.118.263202,PhysRevLett.119.253001,PhysRevLett.121.263202,PhysRevA.87.012509,PhysRevLett.109.263004,2103.12052}. Improved clock stability allows more precise measurements of the energy shift of the clock transition, and this leads to better understandings of systematic uncertainties. So far, these are followed by implementations of more sophisticated ways of cancelling the systematic shifts and uncertainties, contributing to the improvement of the stability of the clocks. 

One of the such systematic shifts is gravitational redshift. The high clock stability has achieved a resolution in the gravitational redshift induced by the Earth's gravity in the lab. The first report had a resolution of tens of centimeters \cite{Science.329.1630}, and a following report has the precision of 1 cm height difference consistent with the convensional geodesy \cite{Nat.Photon.10.662}, and the measurement outside the lab also demonstrated the similar resolution \cite{Nat.Photon.14.411}. A recent report with the ultimate stability of $7.6\times 10^{-21}$ with $10^5$ s integration resolves the gravitational redshift within a vertically-elongated atomic ensemble \cite{2109.12238}. Further improvement of the stability can resolve atoms in different lattice sites in a single Ramsey sequence, leading to gravity's distinguishing internal structure of a coherent spin state (CSS). Together with an entanglement in an atomic ensemble, this leads to the gravity's accessing to a quantum state, which casts a new light to the discussion of gravitational decoherence. In this paper, we first analyze how the decoherence occurs to a CSS, quantify the signal for this in the context of the stability of optical lattice clocks, and then discuss how we should interpret this analysis in the context of gravity's decohering a quantum state. Finally, we discuss the feasibility of the observation of the decoherence with the state-of-the-art optical lattice clocks.

\section{Gravitational redshift in an atomic ensemble}
Suppose an atomic ensemble consisting of multiple atoms is under the Earth's gravitational field. The amount of the gravitational redshift $\Delta \nu$ of an atomic transition of frequency $\nu$ between two places on Earth with a height difference of $\Delta h$ is
\begin{equation}
\frac{\Delta \nu}{\nu}=\frac{g\Delta h}{c^2},
\end{equation}
assuming that the gravitational acceleration $g=9.80665$ m/s$^2$ is locally uniform around the two places, where $c$ is the speed of light. The amount of the relative shift is $\Delta \nu/\nu=1.09\times 10^{-18}$ for $\Delta h=1$ cm, consistent with the 1 cm height resolution for the currently reported best stability around $10^{-18}$ \cite{NatPhoton.9.185,NatCommun.6.6896,Science.341.1215,PhysRevLett.116.063001,Metrologia.56.065004,PhysRevLett.123.033201}. Naively, if the stability increases up to $10^{-23}$, the resolution in the height is $\sim 100 $ nm, which is the same order of magnitude as the height difference between each lattice site. 

The fundamental limit of the stability of the optical lattice clock comes from the quantum projection noise (QPN) at the state detection \cite{PhysRevA.47.3554}. The relative stability of the clock due to the QPN is described as
\begin{equation}\label{Eq:SQL}
\sigma_{\rm QPN}=\frac{1}{\omega_0 \tau_{\rm R}}\sqrt{\frac{T_{\rm C}}{\tau}} \sqrt{\frac{\xi^2_{\rm W}}{N}},
\end{equation}
where $\omega_0$ is the resonant angular frequency of the clock transition, $\tau_{\rm R}$ is the interrogation time, $T_{\rm C}$ is the time for one cycle of the clock operation, $\tau$ is the overall integration time, $\xi^2_{\rm W}$ is the Wineland parameter, and $N$ is the number of atoms in the system \cite{Nature.588.414}. The amount of the QPN for a CSS, where $\xi^2_{\rm W}=1$, is called the standard quantum limit (SQL) and with a squeezed spin state and other entangled states, $\xi^2_{\rm W}$ can be smaller than 1. The smaller SQL is more beneficial to detect the decoherence by gravity. 

Current state-of-the-art optical lattice clocks and clocks based on neutral atoms have $\tau_{\rm R}=30$ s \cite{Nature.588.408,2109.12238,2109.12237}. Because the typical time for loading atoms and post-interrogation measurement is at most 1 s in total, when $\tau_{\rm R}=30$ s, $T_{\rm C}/\tau_R\simeq 1$. To set the distance between the atoms in each lattice site and $\omega_0$, an optical lattice clock with ytterbium (Yb) is assumed, where $\omega_0=2\pi \times 5.18295 \times 10^{14}$ Hz and magic wavelength $\lambda=759.356$ nm \cite{PhysRevLett.119.253001}. 

In the assumed system, single atoms are located at each site of a cubic three-dimensional lattice of $n_{\rm site}$ sites along the $x$ and $y$ axes and $n_{\rm site}+1$ sites along the $z$ axis, resulting in $N=n_{\rm site}^2(n_{\rm site}+1)$. A three-dimensional optical lattice is beneficial to eliminate density shift, and a cube is the rectangular three-dimensional object, which is theoretically easy to deal with in this context, closest to the optical lattice generated by three Gaussian beams. The probe beam is sent along the $z$ direction, which is a typical configuration for optical lattice clocks and also helps reducing the broadening due to tunneling effect \cite{PhysRevA.72.033409}. Because the effect of the gravitational redshift appears along the $z$ axis, it is convenient to assume that $n_{\rm site}^2$ atoms in each layer perpendicular to the $z$ axis forms a single CSS. The CSS is defined as decohered when the phase difference between the top and the bottom layer after the interrogation time is as large as the SQL of each layer. To estimate time evolution of a single quantum state, a single interrogation is considered, resulting in $\tau=\tau_0$. These assumptions reduces Eq. \ref{Eq:SQL} to $\sigma_{\rm QPN}=1/\omega_0 \tau n_{\rm site}$ for each layer, and the amount of the gravitational redshift between the top and the bottom layer is $g n_{\rm site}\lambda/2c^2$. Equating these two gives $n_{\rm site}=497$, leading to $\sigma_{\rm QPN}=2.06\times10^{-20}$ for a single layer and $\sigma_{\rm QPN}=9.23\times10^{-22}$ for whole $N$ atoms. With these conditions, the top layer is incoherent with the bottom layer after the time evelution over 30 s, and the whole atomic ensemble that was initially coherent is no longer a single CSS. 

The total atom number $N=1.23\times10^8$ in the system of $n_{\rm site}=497$ is orders of magnitude larger than the largerst number mentioned in Refs. \cite{Science.358.6359,2109.12238}. Also, the length of the one edge of the cube is similar to the beam size of 170 $\mu$m for a one-dimensional lattice \cite{2109.12238}. From these perspectives, decent amount of technical developement is necessary to observe the gravitational dephasing in the optical lattice clock. However, the dephasing can be described not between different layers but between larger portion of the atomic ensemble. For example, when the coherence is defined between the upper half and the lower half, the decoherence is observable at $n_{\rm site}=165$, corresponding to $N=4.49\times10^6$. This increases the feasibility substantially.

To quantify the effect further, the behavior of the whole system is calculated by summing up that of each layer. The overall atomic ensemble can be described as the sum of the $n_{\rm site}+1$ layers perpendicular to the $z$ axis. Because the relative phase shift of the upper half is symmetric to the lower half except for the sign of it, the phase drift between the atomic system and the local oscillator observed in the Ramsey sequence does not change even with the existence of the significant gravitational redshift. However, relative shift between each layer shortens the effective length of the overall Bloch vector under the gravitational redshift. 

\begin{figure}[!t]
    \includegraphics[width=1\columnwidth]{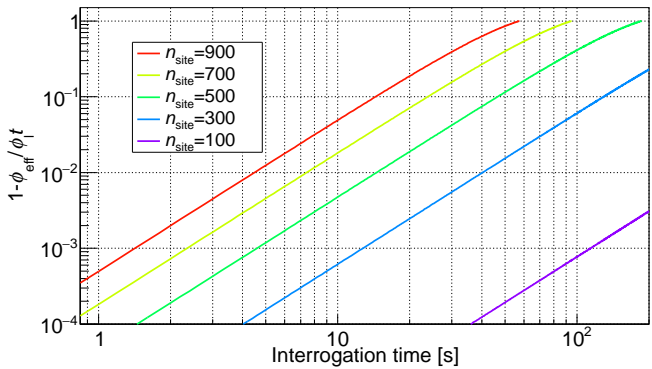}
    \caption{(Color online) Deviation of the ratio of the measured phase to the expected phase $\phi_{\rm eff}/\phi_{\rm l}t$ from 1 versus the interrogation time for different $n_{\rm site}$: $\phi_l$ is assumed to be $10^{-5}$s$^{-1}$. }
    \label{PhaseLoss}
\end{figure}

The shortening of the overall Bloch vector is numerically estimated by the sum of the Bloch vector for each layer:
\begin{equation}
\begin{pmatrix} 
S_{\rm x} \\
S_{\rm y} 
\end{pmatrix}
=\sum^{n_{\rm site}/2}_{k=-n_{\rm site}/2}
\begin{pmatrix} 
\cos(\phi_{\rm l}+k\phi_{\rm g})t \\ 
\sin(\phi_{\rm l}+k\phi_{\rm g})t 
\end{pmatrix}
,
\end{equation}
where $S_{\rm x(y)}$ is the $x$ ($y$) component of the overall Bloch vector, $\phi_{\rm l}$ is the phase shift per unit time due to the phase drift of the laser, and $\phi_{\rm g}$ is the phase drift per unit time due to the gravitational redshift per layer. The length of the Bloch vector for a single layer is normalized to 1. The nominal phase shift due to the laser drift without the gravitational redshift is $\phi_{\rm l} t$, and the measured phase shift $\phi_{\rm eff}$ is calculated as $\phi_{\rm eff}=\sin^{-1}(S_{\rm y}/n_{\rm site}+1)$. Figure \ref{PhaseLoss} shows $\phi_{\rm eff}/\phi_{\rm l} t$ for different $n_{\rm site}$. At $n_{\rm site}=100$, no significant degradation in the phase measurement is observed even for $\tau=200$ s, but for $n_{\rm site}=500$, the measured phase drift underestimates the actual phase drift by 40 \% at $\tau=100$ s. This clearly shows that the stability of the atomic clock is limited by the gravitational dephasing.  

\begin{figure}[!t]
    \includegraphics[width=1\columnwidth]{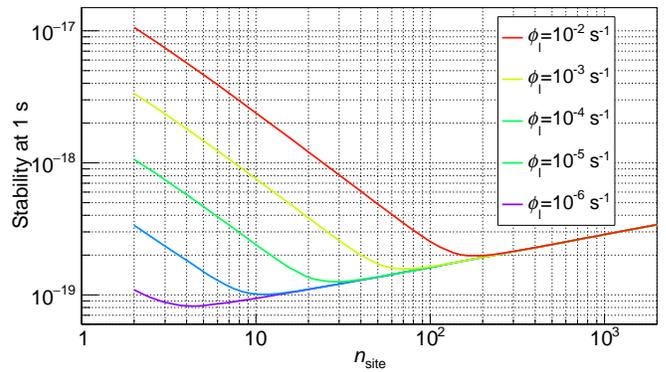}
    \caption{(Color online) Best stability at 1 s achievable for a cubic atomic ensemble with $n_{\rm site}$ atoms along each axis: to compare different interrogation times in the same plot, the stability at $\tau$ is converted to the stability at 1 s integration time by using the $\tau^{-1/2}$ scaling. $\phi_{\rm l}$ of 1 s$^{-1}$ corresponds to the frequency difference of $1/2\pi$ Hz, and thus $\phi_{\rm l}$ shown in the plot corresponds to the linewidth ranging from $0.16$ $\mu$Hz to 1.6 mHz. }
    \label{NsiteVSSQL}
\end{figure}

The maximally allowed interrogation time is defined as the time when the systematic shift due to the decoherence is as large as the SQL, which is $1/n_{\rm site}$ for the Bloch vector for a single layer normalized to 1. Figure \ref{NsiteVSSQL} shows the stability at such maximum interrogation time for different $n_{\rm site}$ and $\phi_{\rm l}$. At small $n_{\rm site}$, the stability roughly scales to $n_{\rm site}^{-1}$. This is when the stability is limited by the coherence of the laser for small $n_{\rm site}$, and $\tau$ is basically more or less constant. In fact, for the smallest $n_{\rm site}=2$ and $\phi_{\rm l}=10^{-6}$ s$^{-1}$, to obtain the stability shown in the plot, $\tau$ has to be as large as $1.97\times10^6$, which satisfies $\phi_{\rm l} \tau \sim 1$. This $\tau$ is unrealistically large given the current technology on the stabilization of the laser and atomic system. Assuming the laser is infinitely stable, the plot is dominated by the $\sim n_{\rm site}^{0.25}$ scaling due to nonzero $\phi_{\rm g}$. The minimum of the stability is determined by the compromise between these two factors. This is, for intstance for $\phi_{\rm l}=10^{-2}$ s$^{-1}$, $2\times10^{-19}$ with $n_{\rm site}\simeq 200$. Because the scaling of the stability to $\phi_{\rm g}$ is not large, regardless of the improvement in the laser stability, the stability of the optical lattice clock is limited to $\sim 10^{-19}$ per 1 s due to the gravitational redshift as far as the distribution of atoms is symmetric for the $x$, $y$, and $z$ axes. Note that this is consistent with the naive estimate that 100 $\mu$m shift of the height can be detected with $10^{-20}$ stability clock; the interrogation time required to reach $2\times10^{-19}$ with $n_{\rm site}= 200$ is $\tau=60$ s, and therefore actual stability in a single sequence is $2.58\times10^{-20}$. This is larger than the stability limit $1.30 \times 10^{-21} $ of the relativistic limit on a trapped single atom discussed in Ref.~\cite{ClassQuantumGrav.32.015018}. 

\begin{figure}[!t]
    \includegraphics[width=1\columnwidth]{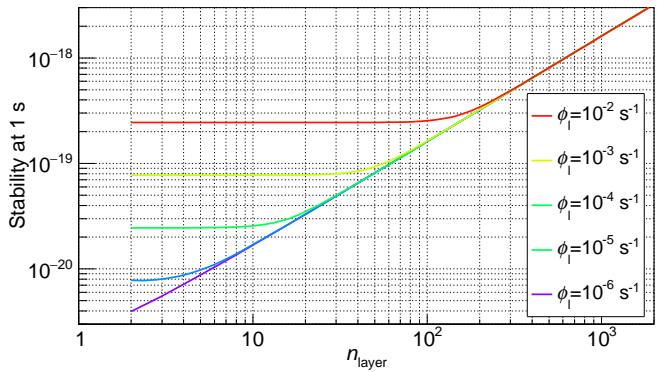}
    \caption{(Color online) Best stability at 1 s achievable for one-dimensional optical lattice with $n_{\rm layer}$: to compare different interrogation times in the same plot, the stability at $\tau$ is converted to the stability at 1 s integration time by using the $\tau^{-1/2}$ scaling. $\phi_{\rm l}$ of 1 s$^{-1}$ corresponds to the frequency difference of $1/2\pi$ Hz, and thus $\phi_{\rm l}$ shown in the plot corresponds to the linewidth ranging from $0.16$ $\mu$Hz to 1.6 mHz. }
    \label{NlayerVSSQL}
\end{figure}

Another geometry that can be easier to realize is one-dimensional lattice whose symmetry axis is along the $z$ axis, which is the same geometry as Ref. \cite{2109.12238}. To simulate this case, nonuniform density shift between layers makes it difficult to systematically subtract the dephasing due to the density shift. An elongated three-dimensional lattice along $z$ axis can eliminate such nonuniformity in the density shift. In this case, the number of atoms in a single layer is fixed at $n_{\rm site}^2=10^4$, and the number of layers $n_{\rm layer}$ is changed. Figure \ref{NlayerVSSQL} is obtained from the same analysis as Fig. \ref{NsiteVSSQL}. The flat part is the stability limited by the phase drift of the laser, and the slope of $({\rm stability~at~1~s})\propto n_{\rm layer}$ is due to the dephasing by gravity, meaning that the stability of the clock is limited by the gravitational dephasing at large $n_{\rm layer}$. On one hand, this means that the gravitational dephasing limits the next generation optical lattice clocks at large $n_{\rm layer}$. On the other hand, this limited stability after removing all the systematic effects is the experimental signal of the gravitational dephasing.

This shows that decoherence of a CSS is induced by an effect of gravity, meaning that a high-precision optical lattice clock can be a tool to observe decoherence of a quantum state induced by gravity for the first time. To guarantee that the decoherence occurring in a optical lattice clock is that of a quantum state, the atomic ensemble needs to be an entangled state of a certain kind, as a CSS is a classical state. An implementation of an entangled state in an optical lattice clock is demonstrated for a spin squeezed state \cite{Nature.588.414}, and more complicated state (e.g., non-Gaussian states \cite{Nature.519.439}) can be implemented in a similar manner. The stability of an optical lattice clock with entangled states is described with Allan deviation, and the signature of the decoherence observable by the clock operation is the limitation in its stability similar to Figs. \ref{NsiteVSSQL} and \ref{NlayerVSSQL}. Naively, Eq. \ref{Eq:SQL} shows the reduction of the QPN by a factor of $\xi_{\rm W}$. In addition, entangled states are more sensitive to the decoherence than a CSS. Unless the inter-layer entanglement is negligible, the dephasing between the layers breaks the coherence, which distorts the distribution of the state on the Bloch sphere. Such a change can be detected by a state tomography. Note that any system specific calculation with $10^6$ atoms is beyond the scope of this paper due to the computational capacity. 

This is a different phenomenon from the conventional gravitational decoherence \cite{PhysRevD.68.085006,ClassQuantumGrav.34.193002}. The decoherence of clocks by the interaction between atoms through gravity \cite{PNAS.114.E2303,PhysRevA.95.052116} is also different perspective from the analysis here. Rather, the form of the decoherence resembles the dephasing in the context of NMR induced by an external gravitational field, and thus this phenomenon is called gravitational dephasing in this article. Still, the decoherence of the entangled state described here is induced by an effect of gravity, and as a broader meaning, this is a gravitational decoherence. 

The suppression of the dephasing is difficult, as far as the optical lattice clock is on Earth, unlike the dephasing due to magnetic field or electric field, which can be compensated by applying a constant bias field. The spin echo sequence retains the coherence, but the cancellation of the phase shift of the laser against the atomic system makes it impractical to implement it to the optical lattice clock system. Compensating the shift by applying external field is conceptually possible, but practically difficult. As far as we consider a single Zeeman sublevel as a transition to be interrogated, quadrupole magnetic field can compensate the gravitational redshift. However, we typically interrogate opposite $m_F$ states as well, which behaves oppositely against the magnetic field. If applying an alternating bias field is possible, cancellation of the gravitational dephasing by external compensation field is possible. Making use of the second order Zeeeman shift or DC Stark shift is another possibility, but generating a shift linear to position can be problematic as the shift is quadratic to the field. A way to circumvent the limitation to the clock stability due to the dephasing by the gravitational redshift is to implement the single-atom-resolved detection, which is developed recently in the field of quantum degenerate gas \cite{Nature.462.74,Nature.467.68}. The state-of-the-art single-atom-resolved detection for 2D optical lattice in a single plane is not ideal for trapping a large number of atoms. Currently available best resolution for the detection of a 3D optical lattice is 1.1 $\mu$m \cite{PhysRevLett.120.103201}. Further improvement in this  down to the level that each layer can be detected separately allows the systematic compensation after the measurement. Note that as far as the relative shift within an atomic ensemble is concerned, the uncertainty in the absolute height from the geodesy and any local deviation of the gravitational acceleration from the standard value, which is in the order of 0.1\%, is not a problem.

\section{Suppression of systematic effects to observe the gravitational dephasing}\label{systematics}
To observe this tiny effect of the gravitational dephasing, other systematic shifts needs to be suppressed. 
First of all, the linewidth of the laser of $\sim 1$ mHz or smaller, which is slightly narrower than currently available most stable laser \cite{PhysRevLett.118.263202}, is required, as Fig. \ref{NsiteVSSQL} shows. 

Some systematic shifts for optical lattice clocks are not problems as far as the goal is to observe the relative frequency shift induced by the gravitational redshift, not the absolute shift from the true value. 
The density shift is one of such systematic shifts, because a three-dimensional lattice with single atom in each site has uniform density. 
The energy shift due to diopole-dipole interaction described in Ref. \cite{PhysRevLett.127.013401} does not affect the current analysis substantially. The energy shifts do not monotonically change along the $z$ axis, and because Yb does not have any Zeeman sublevels with small Clebsch-Gordan coefficient, assuming two level system is good approximation of the multilevel system. The overall shift can be cancelled out as a common-mode shift. The effect of tunneling discussed in Ref. \cite{PhysRevA.72.033409} can be suppressed sufficiently with a good alignment of the probe beam to the $z$ axis in the given configuration. In fact, the recoil of the probe light in any direction is not conspicuous in the one-dimensional optical lattice in Ref. \cite{2109.12238} down to the order of $10^{-21}$. 

The AC Stark shift by the probe beam does not matter, either. Typically, the probe beam has a much larger beam size than the trap beam. In this situation, the position dependence of the intensity for the probe beam is much smaller than that of the trap beam, and thus the effect of the probe AC stark shift is subdominant to that of the lattice AC Stark shift. The background gas collision \cite{Nature.564.87} is also negligible, because it happens uniformly over the atomic ensemble. 

Other shifts needs to be carefully estimated. Here, the assumption is to detect the gravitational redshift with $n_{\rm site}=100$ corresponding to $\Delta z=37.97$~$\mu$m, which gives the frequency difference of $\Delta\nu=2.145\times10^{-6}$~Hz between the top and the bottom layer. 
The allowed field gradient by the coefficient for the first order Zeeman shift of 199.516(2) Hz/G \cite{Nature.564.87} is $2.69 \times 10^{-4}$~G/m. The precise measurement of the local magnetic field down to this level can be performed with narrow-linewidth transitions sensitive to magnetic field. For example, the $^1$S$_0 \rightarrow ^3$P$_2$ transition has the Zeeman splitting of 2.1 MHz/G with 14 s lifetime. The resulting frequency shift of 0.0214 Hz between the top and bottom layers is detectable and therefore proper compensation based on the measurement is possible. To perform such precision spectroscopy with the $^1$S$_0 \rightarrow ^3$P$_2$ transition, it is essential to trap atoms  in the Lamb-Dicke regime with an optical lattice made of a magic wavelength. The magic wavelength for the $^1$S$_0 \rightarrow ^3$P$_2$ transition is theoretically predicted \cite{PhysRevA.98.022501}, and therefore further experimental investigations would lead to the calibration of the magnetic field with the $^1$S$_0 \rightarrow ^3$P$_2$ transition. Also, in the typical operation of the optical lattice clock, the first order Zeeman shift is cancelled by alternatingly measuring the transition frequency of different circular polarizations. Even if the calibration of the magnetic field with the $^1$S$_0 \rightarrow ^3$P$_2$ transition is difficult, similar cancellation can also be applied to the observation of the gravitational redshift. 
With the field gradient limited by the first order Zeeman shift, the second order Zeeman shift of -0.06095(7) Hz/G$^2$ \cite{Nature.564.87} induces a negligible amount of the frequency shift.  
The allowed electric field gradient of $<3.04 \times10^{4}$ (V/m)/m based on the coefficient for the DC Stark shift of $3.626 \times 10^{-6}$ Hz/(V/m)$^2$ \cite{PhysRevLett.108.153002} is achievable by a metallic shield in the vacuum chamber and indium tin oxide coating on the viewports. 

The intensity dependence of the lattice AC Stark shift is quadratic with proper choice of the polarization and frequency, and staying at the proper combination of the intensity and frequency cancels the linear term \cite{PhysRevLett.119.253001}. To estimate the residual shift quadratic to the intensity, the trap beam of $w=170$ $\mu$m \cite{Nature.564.87} is assumed. The ratio of the intensity between the top and bottom layers is proportional to that of the beam area $w^2(z)/w^2(z+100\lambda)$, which has maxima at $z/z_{\rm R}= \pm \sqrt{(100\lambda /z_{\rm})^2+4}/2$ away from the waist, resulting in the maximum intensity change of 0.0846 \%. Because 10 \% intensity change generates $1\times10^{-19}$ shift \cite{PhysRevLett.119.253001}, this is smaller than the gravitational redshift. Proper alignment of the atoms to the waist further reduces the shift. 

Note that the intensity difference of the $x$ and $y$ beams in the $z$ direction due to the Gaussian profile is significantly larger than the shift estimated above for a three-dimensional lattice. However, the profile of the intensity is not monotonic around the center of the beam, and thus the frequency shift does not have the same $z$ position dependence as the gravitational redshift. This provides a way to discriminate the lattice AC Stark shift from the gravitational redshift. The vector AC Stark shift, which matters for the accurate operation of the optical lattice clock, is cancelled out as a common mode background. 


The difficulties in estimating the black body radiation (BBR) shift in the accurate operation of the optical lattice clock is eased to the relative shift between different parts of the atomic ensemble with respect to the gravitational redshift detection. However, the nonuniform temperature of the surrounding structure that can generate the transition frequency shift linear to the position, similar to the gravitational redshift, still needs to be estimated. Suppose the two opposite chamber walls have temperature of $T_1=293$ K and $T_2=294$ K. Based on the typical size of the vacuum chamber for the optical lattice clock, the chamber walls are assumed to be 5 cm away. This gives the BBR shift ratio of
$(T_2^4d\Omega_+T_1^4d\Omega_-)/(T_2^4d\Omega_-+T_1^4d\Omega_+)$ between two ends of the atomic ensemble, where $d\Omega_{+(-)}$ is the solid angle for the nearest (farthest) layer from the wall. This gives $1.04\times10^{-5}$ difference in the BBR field, resulting in $2.46\times10^{-20}$ difference in the BBR shift \cite{Nature.564.87}. To observe the gravitational redshift, uniformity of the temperature over the chamber in the level of 10 mK is desired. Note that this estimate only sets an upper limit for the one hemisphere being $T_1$ and the other being $T_2$, and system-specific estimates can give smaller amount of the diffference in the BBR field, leading to less stringent requirement for the temperature uniformity. Also, designing a vacuum chamber in a way to suppress the temperature nonuniformity in the vertical direction is possible, e.g. by locating the oven at the same height as the atomic ensemble. If such suppressions are not good enough, another transition that has a different sensitivity to the BBR shift can be utilized for an {\it in situ} evaluation of the amount of the BBR shift \cite{IEEETUFFC.65.1069}. 

Overall, the suppression of the systematic backgrounds to the gravitational redshift looks achievable enough to observe the gravitational dephasing.

\section{Conclusion}
The high precision of the optical lattice clock is reaching the level that the decoherence of a coherent spin state by gravity is observable. This can be visible at the relative stability of $10^{-21}$, and would limit the 1 s stability of the clock to $10^{-19}$ as far as the atomic ensemble is symmetric against all three directions. The limitation in the clock stability worse than that due to laser stability can be the first observation of the decoherence of a quantum state induced by gravity when an entangled state is used for the clock operations. The observation is only possible when other systematic shifts are removed, which seems reasonably feasible. 

\section*{Acknowledgment}
A.K. acknowledges the partial support of a William M. and Jane D. Fairbank Postdoctoral Fellowship of Stanford University.

\bibliography{LOACS}

\begin{thebibliography}{45}%
\makeatletter
\providecommand \@ifxundefined [1]{%
 \@ifx{#1\undefined}
}%
\providecommand \@ifnum [1]{%
 \ifnum #1\expandafter \@firstoftwo
 \else \expandafter \@secondoftwo
 \fi
}%
\providecommand \@ifx [1]{%
 \ifx #1\expandafter \@firstoftwo
 \else \expandafter \@secondoftwo
 \fi
}%
\providecommand \natexlab [1]{#1}%
\providecommand \enquote  [1]{``#1''}%
\providecommand \bibnamefont  [1]{#1}%
\providecommand \bibfnamefont [1]{#1}%
\providecommand \citenamefont [1]{#1}%
\providecommand \href@noop [0]{\@secondoftwo}%
\providecommand \href [0]{\begingroup \@sanitize@url \@href}%
\providecommand \@href[1]{\@@startlink{#1}\@@href}%
\providecommand \@@href[1]{\endgroup#1\@@endlink}%
\providecommand \@sanitize@url [0]{\catcode `\\12\catcode `\$12\catcode
  `\&12\catcode `\#12\catcode `\^12\catcode `\_12\catcode `\%12\relax}%
\providecommand \@@startlink[1]{}%
\providecommand \@@endlink[0]{}%
\providecommand \url  [0]{\begingroup\@sanitize@url \@url }%
\providecommand \@url [1]{\endgroup\@href {#1}{\urlprefix }}%
\providecommand \urlprefix  [0]{URL }%
\providecommand \Eprint [0]{\href }%
\providecommand \doibase [0]{https://doi.org/}%
\providecommand \selectlanguage [0]{\@gobble}%
\providecommand \bibinfo  [0]{\@secondoftwo}%
\providecommand \bibfield  [0]{\@secondoftwo}%
\providecommand \translation [1]{[#1]}%
\providecommand \BibitemOpen [0]{}%
\providecommand \bibitemStop [0]{}%
\providecommand \bibitemNoStop [0]{.\EOS\space}%
\providecommand \EOS [0]{\spacefactor3000\relax}%
\providecommand \BibitemShut  [1]{\csname bibitem#1\endcsname}%
\let\auto@bib@innerbib\@empty
\bibitem [{\citenamefont {Di{\'{o}}si}(2011)}]{JPhysConfSer.306.012006}%
  \BibitemOpen
  \bibfield  {author} {\bibinfo {author} {\bibfnamefont {L.}~\bibnamefont
  {Di{\'{o}}si}},\ }\href {https://doi.org/10.1088/1742-6596/306/1/012006}
  {\bibfield  {journal} {\bibinfo  {journal} {Journal of Physics: Conference
  Series}\ }\textbf {\bibinfo {volume} {306}},\ \bibinfo {pages} {012006}
  (\bibinfo {year} {2011})}\BibitemShut {NoStop}%
\bibitem [{\citenamefont {Penrose}(1996)}]{GenRelGrav.28.581}%
  \BibitemOpen
  \bibfield  {author} {\bibinfo {author} {\bibfnamefont {R.}~\bibnamefont
  {Penrose}},\ }\href {https://doi.org/10.1007/BF02105068} {\bibfield
  {journal} {\bibinfo  {journal} {General Relativity and Gravitation}\ }\textbf
  {\bibinfo {volume} {28}},\ \bibinfo {pages} {581} (\bibinfo {year}
  {1996})}\BibitemShut {NoStop}%
\bibitem [{\citenamefont {Diósi}(1987)}]{PhysLettA.120.377}%
  \BibitemOpen
  \bibfield  {author} {\bibinfo {author} {\bibfnamefont {L.}~\bibnamefont
  {Diósi}},\ }\href
  {https://doi.org/https://doi.org/10.1016/0375-9601(87)90681-5} {\bibfield
  {journal} {\bibinfo  {journal} {Physics Letters A}\ }\textbf {\bibinfo
  {volume} {120}},\ \bibinfo {pages} {377} (\bibinfo {year}
  {1987})}\BibitemShut {NoStop}%
\bibitem [{\citenamefont {Bassi}\ \emph {et~al.}(2017)\citenamefont {Bassi},
  \citenamefont {Gro{\ss}ardt},\ and\ \citenamefont
  {Ulbricht}}]{ClassQuantumGrav.34.193002}%
  \BibitemOpen
  \bibfield  {author} {\bibinfo {author} {\bibfnamefont {A.}~\bibnamefont
  {Bassi}}, \bibinfo {author} {\bibfnamefont {A.}~\bibnamefont
  {Gro{\ss}ardt}},\ and\ \bibinfo {author} {\bibfnamefont {H.}~\bibnamefont
  {Ulbricht}},\ }\href {https://doi.org/10.1088/1361-6382/aa864f} {\bibfield
  {journal} {\bibinfo  {journal} {Class. Quantum Grav.}\ }\textbf {\bibinfo
  {volume} {34}},\ \bibinfo {pages} {193002} (\bibinfo {year}
  {2017})}\BibitemShut {NoStop}%
\bibitem [{\citenamefont {Kok}\ and\ \citenamefont
  {Yurtsever}(2003)}]{PhysRevD.68.085006}%
  \BibitemOpen
  \bibfield  {author} {\bibinfo {author} {\bibfnamefont {P.}~\bibnamefont
  {Kok}}\ and\ \bibinfo {author} {\bibfnamefont {U.}~\bibnamefont
  {Yurtsever}},\ }\href {https://doi.org/10.1103/PhysRevD.68.085006} {\bibfield
   {journal} {\bibinfo  {journal} {Phys. Rev. D}\ }\textbf {\bibinfo {volume}
  {68}},\ \bibinfo {pages} {085006} (\bibinfo {year} {2003})}\BibitemShut
  {NoStop}%
\bibitem [{\citenamefont {Khosla}\ and\ \citenamefont
  {Altamirano}(2017)}]{PhysRevA.95.052116}%
  \BibitemOpen
  \bibfield  {author} {\bibinfo {author} {\bibfnamefont {K.~E.}\ \bibnamefont
  {Khosla}}\ and\ \bibinfo {author} {\bibfnamefont {N.}~\bibnamefont
  {Altamirano}},\ }\href {https://doi.org/10.1103/PhysRevA.95.052116}
  {\bibfield  {journal} {\bibinfo  {journal} {Phys. Rev. A}\ }\textbf {\bibinfo
  {volume} {95}},\ \bibinfo {pages} {052116} (\bibinfo {year}
  {2017})}\BibitemShut {NoStop}%
\bibitem [{\citenamefont {Castro~Ruiz}\ \emph {et~al.}(2017)\citenamefont
  {Castro~Ruiz}, \citenamefont {Giacomini},\ and\ \citenamefont
  {Brukner}}]{PNAS.114.E2303}%
  \BibitemOpen
  \bibfield  {author} {\bibinfo {author} {\bibfnamefont {E.}~\bibnamefont
  {Castro~Ruiz}}, \bibinfo {author} {\bibfnamefont {F.}~\bibnamefont
  {Giacomini}},\ and\ \bibinfo {author} {\bibfnamefont {{\v C}.}~\bibnamefont
  {Brukner}},\ }\href {https://doi.org/10.1073/pnas.1616427114} {\bibfield
  {journal} {\bibinfo  {journal} {Proc. Nat. Acad. Sci.}\ }\textbf {\bibinfo
  {volume} {114}},\ \bibinfo {pages} {E2303} (\bibinfo {year}
  {2017})}\BibitemShut {NoStop}%
\bibitem [{\citenamefont {Ludlow}\ \emph {et~al.}(2015)\citenamefont {Ludlow},
  \citenamefont {Boyd}, \citenamefont {Ye}, \citenamefont {Peik},\ and\
  \citenamefont {Schmidt}}]{RevModPhys.87.637}%
  \BibitemOpen
  \bibfield  {author} {\bibinfo {author} {\bibfnamefont {A.~D.}\ \bibnamefont
  {Ludlow}}, \bibinfo {author} {\bibfnamefont {M.~M.}\ \bibnamefont {Boyd}},
  \bibinfo {author} {\bibfnamefont {J.}~\bibnamefont {Ye}}, \bibinfo {author}
  {\bibfnamefont {E.}~\bibnamefont {Peik}},\ and\ \bibinfo {author}
  {\bibfnamefont {P.~O.}\ \bibnamefont {Schmidt}},\ }\href
  {https://doi.org/10.1103/RevModPhys.87.637} {\bibfield  {journal} {\bibinfo
  {journal} {Rev. Mod. Phys.}\ }\textbf {\bibinfo {volume} {87}},\ \bibinfo
  {pages} {637} (\bibinfo {year} {2015})}\BibitemShut {NoStop}%
\bibitem [{\citenamefont {Poli}\ \emph {et~al.}(2013)\citenamefont {Poli},
  \citenamefont {Oates}, \citenamefont {Gill},\ and\ \citenamefont
  {Tino}}]{LaRivNuovoCimento.12.555}%
  \BibitemOpen
  \bibfield  {author} {\bibinfo {author} {\bibfnamefont {N.}~\bibnamefont
  {Poli}}, \bibinfo {author} {\bibfnamefont {C.~W.}\ \bibnamefont {Oates}},
  \bibinfo {author} {\bibfnamefont {P.}~\bibnamefont {Gill}},\ and\ \bibinfo
  {author} {\bibfnamefont {G.~M.}\ \bibnamefont {Tino}},\ }\href
  {https://doi.org/10.1393/ncr/i2013-10095-x} {\bibfield  {journal} {\bibinfo
  {journal} {La Rivista del Nuovo Cimento}\ }\textbf {\bibinfo {volume} {36}},\
  \bibinfo {pages} {555} (\bibinfo {year} {2013})}\BibitemShut {NoStop}%
\bibitem [{\citenamefont {Ushihima}\ \emph {et~al.}(2015)\citenamefont
  {Ushihima}, \citenamefont {Takamoto}, \citenamefont {Das}, \citenamefont
  {Ohkubo},\ and\ \citenamefont {Katori}}]{NatPhoton.9.185}%
  \BibitemOpen
  \bibfield  {author} {\bibinfo {author} {\bibfnamefont {I.}~\bibnamefont
  {Ushihima}}, \bibinfo {author} {\bibfnamefont {M.}~\bibnamefont {Takamoto}},
  \bibinfo {author} {\bibfnamefont {M.}~\bibnamefont {Das}}, \bibinfo {author}
  {\bibfnamefont {T.}~\bibnamefont {Ohkubo}},\ and\ \bibinfo {author}
  {\bibfnamefont {H.}~\bibnamefont {Katori}},\ }\href
  {https://doi.org/https://doi.org/10.1038/nphoton.2015.5} {\bibfield
  {journal} {\bibinfo  {journal} {Nat. Photon.}\ }\textbf {\bibinfo {volume}
  {9}},\ \bibinfo {pages} {185} (\bibinfo {year} {2015})}\BibitemShut {NoStop}%
\bibitem [{\citenamefont {Nicholson}\ \emph {et~al.}(2015)\citenamefont
  {Nicholson}, \citenamefont {Campbell}, \citenamefont {Hutson}, \citenamefont
  {Marti}, \citenamefont {Bloom}, \citenamefont {McNally}, \citenamefont
  {Zhang}, \citenamefont {Barrett}, \citenamefont {Safronova}, \citenamefont
  {Strouse}, \citenamefont {Tew},\ and\ \citenamefont {Ye}}]{NatCommun.6.6896}%
  \BibitemOpen
  \bibfield  {author} {\bibinfo {author} {\bibfnamefont {T.~L.}\ \bibnamefont
  {Nicholson}}, \bibinfo {author} {\bibfnamefont {S.~L.}\ \bibnamefont
  {Campbell}}, \bibinfo {author} {\bibfnamefont {R.~B.}\ \bibnamefont
  {Hutson}}, \bibinfo {author} {\bibfnamefont {G.~E.}\ \bibnamefont {Marti}},
  \bibinfo {author} {\bibfnamefont {B.~J.}\ \bibnamefont {Bloom}}, \bibinfo
  {author} {\bibfnamefont {R.~L.}\ \bibnamefont {McNally}}, \bibinfo {author}
  {\bibfnamefont {W.}~\bibnamefont {Zhang}}, \bibinfo {author} {\bibfnamefont
  {M.~D.}\ \bibnamefont {Barrett}}, \bibinfo {author} {\bibfnamefont {M.~S.}\
  \bibnamefont {Safronova}}, \bibinfo {author} {\bibfnamefont {G.~F.}\
  \bibnamefont {Strouse}}, \bibinfo {author} {\bibfnamefont {W.~L.}\
  \bibnamefont {Tew}},\ and\ \bibinfo {author} {\bibfnamefont {J.}~\bibnamefont
  {Ye}},\ }\href {https://doi.org/https://doi.org/10.1038/ncomms7896}
  {\bibfield  {journal} {\bibinfo  {journal} {Nat. Commun.}\ }\textbf {\bibinfo
  {volume} {6}},\ \bibinfo {pages} {6896} (\bibinfo {year} {2015})}\BibitemShut
  {NoStop}%
\bibitem [{\citenamefont {Hinkley}\ \emph {et~al.}(2013)\citenamefont
  {Hinkley}, \citenamefont {Sherman}, \citenamefont {Phillips}, \citenamefont
  {Schioppo}, \citenamefont {Lemke}, \citenamefont {Beloy}, \citenamefont
  {Pizzocaro}, \citenamefont {Oates},\ and\ \citenamefont
  {Ludlow}}]{Science.341.1215}%
  \BibitemOpen
  \bibfield  {author} {\bibinfo {author} {\bibfnamefont {N.}~\bibnamefont
  {Hinkley}}, \bibinfo {author} {\bibfnamefont {J.~A.}\ \bibnamefont
  {Sherman}}, \bibinfo {author} {\bibfnamefont {N.~B.}\ \bibnamefont
  {Phillips}}, \bibinfo {author} {\bibfnamefont {M.}~\bibnamefont {Schioppo}},
  \bibinfo {author} {\bibfnamefont {N.~D.}\ \bibnamefont {Lemke}}, \bibinfo
  {author} {\bibfnamefont {K.}~\bibnamefont {Beloy}}, \bibinfo {author}
  {\bibfnamefont {M.}~\bibnamefont {Pizzocaro}}, \bibinfo {author}
  {\bibfnamefont {C.~W.}\ \bibnamefont {Oates}},\ and\ \bibinfo {author}
  {\bibfnamefont {A.~D.}\ \bibnamefont {Ludlow}},\ }\href
  {https://doi.org/10.1126/science.1240420} {\bibfield  {journal} {\bibinfo
  {journal} {Science}\ }\textbf {\bibinfo {volume} {341}},\ \bibinfo {pages}
  {1215} (\bibinfo {year} {2013})}\BibitemShut {NoStop}%
\bibitem [{\citenamefont {Huntemann}\ \emph {et~al.}(2016)\citenamefont
  {Huntemann}, \citenamefont {Sanner}, \citenamefont {Lipphardt}, \citenamefont
  {Tamm},\ and\ \citenamefont {Peik}}]{PhysRevLett.116.063001}%
  \BibitemOpen
  \bibfield  {author} {\bibinfo {author} {\bibfnamefont {N.}~\bibnamefont
  {Huntemann}}, \bibinfo {author} {\bibfnamefont {C.}~\bibnamefont {Sanner}},
  \bibinfo {author} {\bibfnamefont {B.}~\bibnamefont {Lipphardt}}, \bibinfo
  {author} {\bibfnamefont {C.}~\bibnamefont {Tamm}},\ and\ \bibinfo {author}
  {\bibfnamefont {E.}~\bibnamefont {Peik}},\ }\href
  {https://doi.org/10.1103/PhysRevLett.116.063001} {\bibfield  {journal}
  {\bibinfo  {journal} {Phys. Rev. Lett.}\ }\textbf {\bibinfo {volume} {116}},\
  \bibinfo {pages} {063001} (\bibinfo {year} {2016})}\BibitemShut {NoStop}%
\bibitem [{\citenamefont {Bothwell}\ \emph {et~al.}(2019)\citenamefont
  {Bothwell}, \citenamefont {Kedar}, \citenamefont {Oelker}, \citenamefont
  {Robinson}, \citenamefont {Bromley}, \citenamefont {Tew}, \citenamefont
  {Ye},\ and\ \citenamefont {Kennedy}}]{Metrologia.56.065004}%
  \BibitemOpen
  \bibfield  {author} {\bibinfo {author} {\bibfnamefont {T.}~\bibnamefont
  {Bothwell}}, \bibinfo {author} {\bibfnamefont {D.}~\bibnamefont {Kedar}},
  \bibinfo {author} {\bibfnamefont {E.}~\bibnamefont {Oelker}}, \bibinfo
  {author} {\bibfnamefont {J.~M.}\ \bibnamefont {Robinson}}, \bibinfo {author}
  {\bibfnamefont {S.~L.}\ \bibnamefont {Bromley}}, \bibinfo {author}
  {\bibfnamefont {W.~L.}\ \bibnamefont {Tew}}, \bibinfo {author} {\bibfnamefont
  {J.}~\bibnamefont {Ye}},\ and\ \bibinfo {author} {\bibfnamefont {C.~J.}\
  \bibnamefont {Kennedy}},\ }\href {https://doi.org/10.1088/1681-7575/ab4089}
  {\bibfield  {journal} {\bibinfo  {journal} {Metrologia}\ }\textbf {\bibinfo
  {volume} {56}},\ \bibinfo {pages} {065004} (\bibinfo {year}
  {2019})}\BibitemShut {NoStop}%
\bibitem [{\citenamefont {Young}\ \emph {et~al.}(2020)\citenamefont {Young},
  \citenamefont {Eckner}, \citenamefont {J.~Milner}, \citenamefont {Keder},
  \citenamefont {Norcia}, \citenamefont {Oelker}, \citenamefont {Schine},
  \citenamefont {Ye},\ and\ \citenamefont {Kaufman}}]{Nature.588.408}%
  \BibitemOpen
  \bibfield  {author} {\bibinfo {author} {\bibfnamefont {A.~W.}\ \bibnamefont
  {Young}}, \bibinfo {author} {\bibfnamefont {W.}~\bibnamefont {Eckner}},
  \bibinfo {author} {\bibfnamefont {W.~R.}\ \bibnamefont {J.~Milner}}, \bibinfo
  {author} {\bibfnamefont {D.}~\bibnamefont {Keder}}, \bibinfo {author}
  {\bibfnamefont {M.~A.}\ \bibnamefont {Norcia}}, \bibinfo {author}
  {\bibfnamefont {E.}~\bibnamefont {Oelker}}, \bibinfo {author} {\bibfnamefont
  {N.}~\bibnamefont {Schine}}, \bibinfo {author} {\bibfnamefont
  {J.}~\bibnamefont {Ye}},\ and\ \bibinfo {author} {\bibfnamefont {A.~M.}\
  \bibnamefont {Kaufman}},\ }\href
  {https://doi.org/https://doi.org/10.1038/s41586-020-3009-y} {\bibfield
  {journal} {\bibinfo  {journal} {Nature}\ }\textbf {\bibinfo {volume} {588}},\
  \bibinfo {pages} {408} (\bibinfo {year} {2020})}\BibitemShut {NoStop}%
\bibitem [{\citenamefont {Oelker}\ \emph {et~al.}(2019)\citenamefont {Oelker},
  \citenamefont {Hutson}, \citenamefont {Kennedy}, \citenamefont {Sonderhouse},
  \citenamefont {Bothwell}, \citenamefont {Goban}, \citenamefont {Kedar},
  \citenamefont {Sanner}, \citenamefont {Robinson}, \citenamefont {Marti},
  \citenamefont {Legero}, \citenamefont {Giunta}, \citenamefont {Holzwarth},
  \citenamefont {Reihle}, \citenamefont {Sterr},\ and\ \citenamefont
  {Ye}}]{NatPhoton.13.714}%
  \BibitemOpen
  \bibfield  {author} {\bibinfo {author} {\bibfnamefont {E.}~\bibnamefont
  {Oelker}}, \bibinfo {author} {\bibfnamefont {R.~B.}\ \bibnamefont {Hutson}},
  \bibinfo {author} {\bibfnamefont {C.~J.}\ \bibnamefont {Kennedy}}, \bibinfo
  {author} {\bibfnamefont {L.}~\bibnamefont {Sonderhouse}}, \bibinfo {author}
  {\bibfnamefont {T.}~\bibnamefont {Bothwell}}, \bibinfo {author}
  {\bibfnamefont {A.}~\bibnamefont {Goban}}, \bibinfo {author} {\bibfnamefont
  {D.}~\bibnamefont {Kedar}}, \bibinfo {author} {\bibfnamefont
  {C.}~\bibnamefont {Sanner}}, \bibinfo {author} {\bibfnamefont {J.~M.}\
  \bibnamefont {Robinson}}, \bibinfo {author} {\bibfnamefont {G.~E.}\
  \bibnamefont {Marti}}, \bibinfo {author} {\bibfnamefont {T.}~\bibnamefont
  {Legero}}, \bibinfo {author} {\bibfnamefont {M.}~\bibnamefont {Giunta}},
  \bibinfo {author} {\bibfnamefont {R.}~\bibnamefont {Holzwarth}}, \bibinfo
  {author} {\bibfnamefont {F.}~\bibnamefont {Reihle}}, \bibinfo {author}
  {\bibfnamefont {U.}~\bibnamefont {Sterr}},\ and\ \bibinfo {author}
  {\bibfnamefont {J.}~\bibnamefont {Ye}},\ }\href
  {https://doi.org/https://doi.org/10.1038/s41566-019-0493-4} {\bibfield
  {journal} {\bibinfo  {journal} {Nat. Photon.}\ }\textbf {\bibinfo {volume}
  {13}},\ \bibinfo {pages} {714} (\bibinfo {year} {2019})}\BibitemShut
  {NoStop}%
\bibitem [{\citenamefont {McGrew}\ \emph {et~al.}(2018)\citenamefont {McGrew},
  \citenamefont {Zhang}, \citenamefont {Fasano}, \citenamefont {Sch\"affer},
  \citenamefont {Beloy}, \citenamefont {Nicolodi}, \citenamefont {Brown},
  \citenamefont {Hinkley}, \citenamefont {Milani}, \citenamefont {Schioppo},
  \citenamefont {Yoon},\ and\ \citenamefont {Ludlow}}]{Nature.564.87}%
  \BibitemOpen
  \bibfield  {author} {\bibinfo {author} {\bibfnamefont {W.~F.}\ \bibnamefont
  {McGrew}}, \bibinfo {author} {\bibfnamefont {X.}~\bibnamefont {Zhang}},
  \bibinfo {author} {\bibfnamefont {R.~J.}\ \bibnamefont {Fasano}}, \bibinfo
  {author} {\bibfnamefont {S.~A.}\ \bibnamefont {Sch\"affer}}, \bibinfo
  {author} {\bibfnamefont {K.}~\bibnamefont {Beloy}}, \bibinfo {author}
  {\bibfnamefont {D.}~\bibnamefont {Nicolodi}}, \bibinfo {author}
  {\bibfnamefont {R.~C.}\ \bibnamefont {Brown}}, \bibinfo {author}
  {\bibfnamefont {N.}~\bibnamefont {Hinkley}}, \bibinfo {author} {\bibfnamefont
  {G.}~\bibnamefont {Milani}}, \bibinfo {author} {\bibfnamefont
  {M.}~\bibnamefont {Schioppo}}, \bibinfo {author} {\bibfnamefont {T.~H.}\
  \bibnamefont {Yoon}},\ and\ \bibinfo {author} {\bibfnamefont {A.~D.}\
  \bibnamefont {Ludlow}},\ }\href
  {https://doi.org/https://doi.org/10.1038/s41586-018-0738-2} {\bibfield
  {journal} {\bibinfo  {journal} {Nature}\ }\textbf {\bibinfo {volume} {564}},\
  \bibinfo {pages} {87} (\bibinfo {year} {2018})}\BibitemShut {NoStop}%
\bibitem [{\citenamefont {Brewer}\ \emph {et~al.}(2019)\citenamefont {Brewer},
  \citenamefont {Chen}, \citenamefont {Hankin}, \citenamefont {Clements},
  \citenamefont {Chou}, \citenamefont {Wineland}, \citenamefont {Hume},\ and\
  \citenamefont {Leibrandt}}]{PhysRevLett.123.033201}%
  \BibitemOpen
  \bibfield  {author} {\bibinfo {author} {\bibfnamefont {S.~M.}\ \bibnamefont
  {Brewer}}, \bibinfo {author} {\bibfnamefont {J.-S.}\ \bibnamefont {Chen}},
  \bibinfo {author} {\bibfnamefont {A.~M.}\ \bibnamefont {Hankin}}, \bibinfo
  {author} {\bibfnamefont {E.~R.}\ \bibnamefont {Clements}}, \bibinfo {author}
  {\bibfnamefont {C.~W.}\ \bibnamefont {Chou}}, \bibinfo {author}
  {\bibfnamefont {D.~J.}\ \bibnamefont {Wineland}}, \bibinfo {author}
  {\bibfnamefont {D.~B.}\ \bibnamefont {Hume}},\ and\ \bibinfo {author}
  {\bibfnamefont {D.~R.}\ \bibnamefont {Leibrandt}},\ }\href
  {https://doi.org/10.1103/PhysRevLett.123.033201} {\bibfield  {journal}
  {\bibinfo  {journal} {Phys. Rev. Lett.}\ }\textbf {\bibinfo {volume} {123}},\
  \bibinfo {pages} {033201} (\bibinfo {year} {2019})}\BibitemShut {NoStop}%
\bibitem [{\citenamefont {Zheng}\ \emph {et~al.}(2021)\citenamefont {Zheng},
  \citenamefont {Dolde}, \citenamefont {Lochab}, \citenamefont {Merriman},
  \citenamefont {Li},\ and\ \citenamefont {Kolkowitz}}]{2109.12237}%
  \BibitemOpen
  \bibfield  {author} {\bibinfo {author} {\bibfnamefont {X.}~\bibnamefont
  {Zheng}}, \bibinfo {author} {\bibfnamefont {J.}~\bibnamefont {Dolde}},
  \bibinfo {author} {\bibfnamefont {V.}~\bibnamefont {Lochab}}, \bibinfo
  {author} {\bibfnamefont {B.~N.}\ \bibnamefont {Merriman}}, \bibinfo {author}
  {\bibfnamefont {H.}~\bibnamefont {Li}},\ and\ \bibinfo {author}
  {\bibfnamefont {S.}~\bibnamefont {Kolkowitz}},\ }\href@noop {} {\bibfield
  {journal} {\bibinfo  {journal} {arXiv:2109.12237}\ } (\bibinfo {year}
  {2021})},\ \Eprint {https://arxiv.org/abs/2109.12237} {arXiv:2109.12237
  [physics.atom-ph]} \BibitemShut {NoStop}%
\bibitem [{\citenamefont {Bothwell}\ \emph {et~al.}(2021)\citenamefont
  {Bothwell}, \citenamefont {Kennedy}, \citenamefont {Aeppli}, \citenamefont
  {Kedar}, \citenamefont {Robinson}, \citenamefont {Oelker}, \citenamefont
  {Staron},\ and\ \citenamefont {Ye}}]{2109.12238}%
  \BibitemOpen
  \bibfield  {author} {\bibinfo {author} {\bibfnamefont {T.}~\bibnamefont
  {Bothwell}}, \bibinfo {author} {\bibfnamefont {C.~J.}\ \bibnamefont
  {Kennedy}}, \bibinfo {author} {\bibfnamefont {A.}~\bibnamefont {Aeppli}},
  \bibinfo {author} {\bibfnamefont {D.}~\bibnamefont {Kedar}}, \bibinfo
  {author} {\bibfnamefont {J.~M.}\ \bibnamefont {Robinson}}, \bibinfo {author}
  {\bibfnamefont {E.}~\bibnamefont {Oelker}}, \bibinfo {author} {\bibfnamefont
  {A.}~\bibnamefont {Staron}},\ and\ \bibinfo {author} {\bibfnamefont
  {J.}~\bibnamefont {Ye}},\ }\href@noop {} {\bibfield  {journal} {\bibinfo
  {journal} {arXiv:2109.12238}\ } (\bibinfo {year} {2021})},\ \Eprint
  {https://arxiv.org/abs/2109.12238} {arXiv:2109.12238 [physics.atom-ph]}
  \BibitemShut {NoStop}%
\bibitem [{\citenamefont {Pedrozo-Pe\~nafiel}\ \emph
  {et~al.}(2020)\citenamefont {Pedrozo-Pe\~nafiel}, \citenamefont {Colombo},
  \citenamefont {Shu}, \citenamefont {Adiyatullin}, \citenamefont {Li},
  \citenamefont {Mendez}, \citenamefont {Braverman}, \citenamefont {Kawasaki},
  \citenamefont {Akamatsu}, \citenamefont {Xiao},\ and\ \citenamefont
  {Vuleti\'c}}]{Nature.588.414}%
  \BibitemOpen
  \bibfield  {author} {\bibinfo {author} {\bibfnamefont {E.}~\bibnamefont
  {Pedrozo-Pe\~nafiel}}, \bibinfo {author} {\bibfnamefont {S.}~\bibnamefont
  {Colombo}}, \bibinfo {author} {\bibfnamefont {C.}~\bibnamefont {Shu}},
  \bibinfo {author} {\bibfnamefont {A.~F.}\ \bibnamefont {Adiyatullin}},
  \bibinfo {author} {\bibfnamefont {Z.}~\bibnamefont {Li}}, \bibinfo {author}
  {\bibfnamefont {E.}~\bibnamefont {Mendez}}, \bibinfo {author} {\bibfnamefont
  {B.}~\bibnamefont {Braverman}}, \bibinfo {author} {\bibfnamefont
  {A.}~\bibnamefont {Kawasaki}}, \bibinfo {author} {\bibfnamefont
  {D.}~\bibnamefont {Akamatsu}}, \bibinfo {author} {\bibfnamefont
  {Y.}~\bibnamefont {Xiao}},\ and\ \bibinfo {author} {\bibfnamefont
  {V.}~\bibnamefont {Vuleti\'c}},\ }\href
  {https://doi.org/https://doi.org/10.1038/s41586-020-3006-1} {\bibfield
  {journal} {\bibinfo  {journal} {Nature}\ }\textbf {\bibinfo {volume} {588}},\
  \bibinfo {pages} {414} (\bibinfo {year} {2020})}\BibitemShut {NoStop}%
\bibitem [{\citenamefont {Campbell}\ \emph {et~al.}(2017)\citenamefont
  {Campbell}, \citenamefont {Hutson}, \citenamefont {Marti}, \citenamefont
  {Goban}, \citenamefont {Darkwah~Oppong}, \citenamefont {McNally},
  \citenamefont {Sonderhouse}, \citenamefont {Robinson}, \citenamefont {Zhang},
  \citenamefont {Bloom},\ and\ \citenamefont {Ye}}]{Science.358.6359}%
  \BibitemOpen
  \bibfield  {author} {\bibinfo {author} {\bibfnamefont {S.~L.}\ \bibnamefont
  {Campbell}}, \bibinfo {author} {\bibfnamefont {R.~B.}\ \bibnamefont
  {Hutson}}, \bibinfo {author} {\bibfnamefont {G.~E.}\ \bibnamefont {Marti}},
  \bibinfo {author} {\bibfnamefont {A.}~\bibnamefont {Goban}}, \bibinfo
  {author} {\bibfnamefont {N.}~\bibnamefont {Darkwah~Oppong}}, \bibinfo
  {author} {\bibfnamefont {R.~L.}\ \bibnamefont {McNally}}, \bibinfo {author}
  {\bibfnamefont {L.}~\bibnamefont {Sonderhouse}}, \bibinfo {author}
  {\bibfnamefont {J.~M.}\ \bibnamefont {Robinson}}, \bibinfo {author}
  {\bibfnamefont {W.}~\bibnamefont {Zhang}}, \bibinfo {author} {\bibfnamefont
  {B.~J.}\ \bibnamefont {Bloom}},\ and\ \bibinfo {author} {\bibfnamefont
  {J.}~\bibnamefont {Ye}},\ }\href {https://doi.org/10.1126/science.aam5538}
  {\bibfield  {journal} {\bibinfo  {journal} {Science}\ }\textbf {\bibinfo
  {volume} {358}},\ \bibinfo {pages} {90} (\bibinfo {year} {2017})}\BibitemShut
  {NoStop}%
\bibitem [{\citenamefont {Madjarov}\ \emph {et~al.}(2019)\citenamefont
  {Madjarov}, \citenamefont {Cooper}, \citenamefont {Shaw}, \citenamefont
  {Covey}, \citenamefont {Schkolnik}, \citenamefont {Yoon}, \citenamefont
  {Williams},\ and\ \citenamefont {Endres}}]{PhysRevX.9.041052}%
  \BibitemOpen
  \bibfield  {author} {\bibinfo {author} {\bibfnamefont {I.~S.}\ \bibnamefont
  {Madjarov}}, \bibinfo {author} {\bibfnamefont {A.}~\bibnamefont {Cooper}},
  \bibinfo {author} {\bibfnamefont {A.~L.}\ \bibnamefont {Shaw}}, \bibinfo
  {author} {\bibfnamefont {J.~P.}\ \bibnamefont {Covey}}, \bibinfo {author}
  {\bibfnamefont {V.}~\bibnamefont {Schkolnik}}, \bibinfo {author}
  {\bibfnamefont {T.~H.}\ \bibnamefont {Yoon}}, \bibinfo {author}
  {\bibfnamefont {J.~R.}\ \bibnamefont {Williams}},\ and\ \bibinfo {author}
  {\bibfnamefont {M.}~\bibnamefont {Endres}},\ }\href
  {https://doi.org/10.1103/PhysRevX.9.041052} {\bibfield  {journal} {\bibinfo
  {journal} {Phys. Rev. X}\ }\textbf {\bibinfo {volume} {9}},\ \bibinfo {pages}
  {041052} (\bibinfo {year} {2019})}\BibitemShut {NoStop}%
\bibitem [{\citenamefont {Marti}\ \emph {et~al.}(2018)\citenamefont {Marti},
  \citenamefont {Hutson}, \citenamefont {Goban}, \citenamefont {Campbell},
  \citenamefont {Poli},\ and\ \citenamefont {Ye}}]{PhysRevLett.120.103201}%
  \BibitemOpen
  \bibfield  {author} {\bibinfo {author} {\bibfnamefont {G.~E.}\ \bibnamefont
  {Marti}}, \bibinfo {author} {\bibfnamefont {R.~B.}\ \bibnamefont {Hutson}},
  \bibinfo {author} {\bibfnamefont {A.}~\bibnamefont {Goban}}, \bibinfo
  {author} {\bibfnamefont {S.~L.}\ \bibnamefont {Campbell}}, \bibinfo {author}
  {\bibfnamefont {N.}~\bibnamefont {Poli}},\ and\ \bibinfo {author}
  {\bibfnamefont {J.}~\bibnamefont {Ye}},\ }\href
  {https://doi.org/10.1103/PhysRevLett.120.103201} {\bibfield  {journal}
  {\bibinfo  {journal} {Phys. Rev. Lett.}\ }\textbf {\bibinfo {volume} {120}},\
  \bibinfo {pages} {103201} (\bibinfo {year} {2018})}\BibitemShut {NoStop}%
\bibitem [{\citenamefont {Akamatsu}\ \emph {et~al.}(2018)\citenamefont
  {Akamatsu}, \citenamefont {Kobayashi}, \citenamefont {Hisai}, \citenamefont
  {Tanabe}, \citenamefont {Hosaka}, \citenamefont {Yasuda},\ and\ \citenamefont
  {Hong}}]{IEEETUFFC.65.1069}%
  \BibitemOpen
  \bibfield  {author} {\bibinfo {author} {\bibfnamefont {D.}~\bibnamefont
  {Akamatsu}}, \bibinfo {author} {\bibfnamefont {T.}~\bibnamefont {Kobayashi}},
  \bibinfo {author} {\bibfnamefont {Y.}~\bibnamefont {Hisai}}, \bibinfo
  {author} {\bibfnamefont {T.}~\bibnamefont {Tanabe}}, \bibinfo {author}
  {\bibfnamefont {K.}~\bibnamefont {Hosaka}}, \bibinfo {author} {\bibfnamefont
  {M.}~\bibnamefont {Yasuda}},\ and\ \bibinfo {author} {\bibfnamefont {F.-L.}\
  \bibnamefont {Hong}},\ }\href {https://doi.org/10.1109/TUFFC.2018.2819888}
  {\bibfield  {journal} {\bibinfo  {journal} {IEEE Trans. Ultrason.,
  Ferro-electr., Freq. Control}\ }\textbf {\bibinfo {volume} {65}},\ \bibinfo
  {pages} {1069} (\bibinfo {year} {2018})}\BibitemShut {NoStop}%
\bibitem [{\citenamefont {Beloy}\ \emph {et~al.}(2014)\citenamefont {Beloy},
  \citenamefont {Hinkley}, \citenamefont {Phillips}, \citenamefont {Sherman},
  \citenamefont {Schioppo}, \citenamefont {Lehman}, \citenamefont {Feldman},
  \citenamefont {Hanssen}, \citenamefont {Oates},\ and\ \citenamefont
  {Ludlow}}]{PhysRevLett.113.260801}%
  \BibitemOpen
  \bibfield  {author} {\bibinfo {author} {\bibfnamefont {K.}~\bibnamefont
  {Beloy}}, \bibinfo {author} {\bibfnamefont {N.}~\bibnamefont {Hinkley}},
  \bibinfo {author} {\bibfnamefont {N.~B.}\ \bibnamefont {Phillips}}, \bibinfo
  {author} {\bibfnamefont {J.~A.}\ \bibnamefont {Sherman}}, \bibinfo {author}
  {\bibfnamefont {M.}~\bibnamefont {Schioppo}}, \bibinfo {author}
  {\bibfnamefont {J.}~\bibnamefont {Lehman}}, \bibinfo {author} {\bibfnamefont
  {A.}~\bibnamefont {Feldman}}, \bibinfo {author} {\bibfnamefont {L.~M.}\
  \bibnamefont {Hanssen}}, \bibinfo {author} {\bibfnamefont {C.~W.}\
  \bibnamefont {Oates}},\ and\ \bibinfo {author} {\bibfnamefont {A.~D.}\
  \bibnamefont {Ludlow}},\ }\href
  {https://doi.org/10.1103/PhysRevLett.113.260801} {\bibfield  {journal}
  {\bibinfo  {journal} {Phys. Rev. Lett.}\ }\textbf {\bibinfo {volume} {113}},\
  \bibinfo {pages} {260801} (\bibinfo {year} {2014})}\BibitemShut {NoStop}%
\bibitem [{\citenamefont {Beloy}\ \emph {et~al.}(2018)\citenamefont {Beloy},
  \citenamefont {Zhang}, \citenamefont {McGrew}, \citenamefont {Hinkley},
  \citenamefont {Yoon}, \citenamefont {Nicolodi}, \citenamefont {Fasano},
  \citenamefont {Sch\"affer}, \citenamefont {Brown},\ and\ \citenamefont
  {Ludlow}}]{PhysRevLett.120.183201}%
  \BibitemOpen
  \bibfield  {author} {\bibinfo {author} {\bibfnamefont {K.}~\bibnamefont
  {Beloy}}, \bibinfo {author} {\bibfnamefont {X.}~\bibnamefont {Zhang}},
  \bibinfo {author} {\bibfnamefont {W.~F.}\ \bibnamefont {McGrew}}, \bibinfo
  {author} {\bibfnamefont {N.}~\bibnamefont {Hinkley}}, \bibinfo {author}
  {\bibfnamefont {T.~H.}\ \bibnamefont {Yoon}}, \bibinfo {author}
  {\bibfnamefont {D.}~\bibnamefont {Nicolodi}}, \bibinfo {author}
  {\bibfnamefont {R.~J.}\ \bibnamefont {Fasano}}, \bibinfo {author}
  {\bibfnamefont {S.~A.}\ \bibnamefont {Sch\"affer}}, \bibinfo {author}
  {\bibfnamefont {R.~C.}\ \bibnamefont {Brown}},\ and\ \bibinfo {author}
  {\bibfnamefont {A.~D.}\ \bibnamefont {Ludlow}},\ }\href
  {https://doi.org/10.1103/PhysRevLett.120.183201} {\bibfield  {journal}
  {\bibinfo  {journal} {Phys. Rev. Lett.}\ }\textbf {\bibinfo {volume} {120}},\
  \bibinfo {pages} {183201} (\bibinfo {year} {2018})}\BibitemShut {NoStop}%
\bibitem [{\citenamefont {Matei}\ \emph {et~al.}(2017)\citenamefont {Matei},
  \citenamefont {Legero}, \citenamefont {H\"afner}, \citenamefont {Grebing},
  \citenamefont {Weyrich}, \citenamefont {Zhang}, \citenamefont {Sonderhouse},
  \citenamefont {Robinson}, \citenamefont {Ye}, \citenamefont {Riehle},\ and\
  \citenamefont {Sterr}}]{PhysRevLett.118.263202}%
  \BibitemOpen
  \bibfield  {author} {\bibinfo {author} {\bibfnamefont {D.~G.}\ \bibnamefont
  {Matei}}, \bibinfo {author} {\bibfnamefont {T.}~\bibnamefont {Legero}},
  \bibinfo {author} {\bibfnamefont {S.}~\bibnamefont {H\"afner}}, \bibinfo
  {author} {\bibfnamefont {C.}~\bibnamefont {Grebing}}, \bibinfo {author}
  {\bibfnamefont {R.}~\bibnamefont {Weyrich}}, \bibinfo {author} {\bibfnamefont
  {W.}~\bibnamefont {Zhang}}, \bibinfo {author} {\bibfnamefont
  {L.}~\bibnamefont {Sonderhouse}}, \bibinfo {author} {\bibfnamefont {J.~M.}\
  \bibnamefont {Robinson}}, \bibinfo {author} {\bibfnamefont {J.}~\bibnamefont
  {Ye}}, \bibinfo {author} {\bibfnamefont {F.}~\bibnamefont {Riehle}},\ and\
  \bibinfo {author} {\bibfnamefont {U.}~\bibnamefont {Sterr}},\ }\href
  {https://doi.org/10.1103/PhysRevLett.118.263202} {\bibfield  {journal}
  {\bibinfo  {journal} {Phys. Rev. Lett.}\ }\textbf {\bibinfo {volume} {118}},\
  \bibinfo {pages} {263202} (\bibinfo {year} {2017})}\BibitemShut {NoStop}%
\bibitem [{\citenamefont {Brown}\ \emph {et~al.}(2017)\citenamefont {Brown},
  \citenamefont {Phillips}, \citenamefont {Beloy}, \citenamefont {McGrew},
  \citenamefont {Schioppo}, \citenamefont {Fasano}, \citenamefont {Milani},
  \citenamefont {Zhang}, \citenamefont {Hinkley}, \citenamefont {Leopardi},
  \citenamefont {Yoon}, \citenamefont {Nicolodi}, \citenamefont {Fortier},\
  and\ \citenamefont {Ludlow}}]{PhysRevLett.119.253001}%
  \BibitemOpen
  \bibfield  {author} {\bibinfo {author} {\bibfnamefont {R.~C.}\ \bibnamefont
  {Brown}}, \bibinfo {author} {\bibfnamefont {N.~B.}\ \bibnamefont {Phillips}},
  \bibinfo {author} {\bibfnamefont {K.}~\bibnamefont {Beloy}}, \bibinfo
  {author} {\bibfnamefont {W.~F.}\ \bibnamefont {McGrew}}, \bibinfo {author}
  {\bibfnamefont {M.}~\bibnamefont {Schioppo}}, \bibinfo {author}
  {\bibfnamefont {R.~J.}\ \bibnamefont {Fasano}}, \bibinfo {author}
  {\bibfnamefont {G.}~\bibnamefont {Milani}}, \bibinfo {author} {\bibfnamefont
  {X.}~\bibnamefont {Zhang}}, \bibinfo {author} {\bibfnamefont
  {N.}~\bibnamefont {Hinkley}}, \bibinfo {author} {\bibfnamefont
  {H.}~\bibnamefont {Leopardi}}, \bibinfo {author} {\bibfnamefont {T.~H.}\
  \bibnamefont {Yoon}}, \bibinfo {author} {\bibfnamefont {D.}~\bibnamefont
  {Nicolodi}}, \bibinfo {author} {\bibfnamefont {T.~M.}\ \bibnamefont
  {Fortier}},\ and\ \bibinfo {author} {\bibfnamefont {A.~D.}\ \bibnamefont
  {Ludlow}},\ }\href {https://doi.org/10.1103/PhysRevLett.119.253001}
  {\bibfield  {journal} {\bibinfo  {journal} {Phys. Rev. Lett.}\ }\textbf
  {\bibinfo {volume} {119}},\ \bibinfo {pages} {253001} (\bibinfo {year}
  {2017})}\BibitemShut {NoStop}%
\bibitem [{\citenamefont {Ushijima}\ \emph {et~al.}(2018)\citenamefont
  {Ushijima}, \citenamefont {Takamoto},\ and\ \citenamefont
  {Katori}}]{PhysRevLett.121.263202}%
  \BibitemOpen
  \bibfield  {author} {\bibinfo {author} {\bibfnamefont {I.}~\bibnamefont
  {Ushijima}}, \bibinfo {author} {\bibfnamefont {M.}~\bibnamefont {Takamoto}},\
  and\ \bibinfo {author} {\bibfnamefont {H.}~\bibnamefont {Katori}},\ }\href
  {https://doi.org/10.1103/PhysRevLett.121.263202} {\bibfield  {journal}
  {\bibinfo  {journal} {Phys. Rev. Lett.}\ }\textbf {\bibinfo {volume} {121}},\
  \bibinfo {pages} {263202} (\bibinfo {year} {2018})}\BibitemShut {NoStop}%
\bibitem [{\citenamefont {Safronova}\ \emph {et~al.}(2013)\citenamefont
  {Safronova}, \citenamefont {Porsev}, \citenamefont {Safronova}, \citenamefont
  {Kozlov},\ and\ \citenamefont {Clark}}]{PhysRevA.87.012509}%
  \BibitemOpen
  \bibfield  {author} {\bibinfo {author} {\bibfnamefont {M.~S.}\ \bibnamefont
  {Safronova}}, \bibinfo {author} {\bibfnamefont {S.~G.}\ \bibnamefont
  {Porsev}}, \bibinfo {author} {\bibfnamefont {U.~I.}\ \bibnamefont
  {Safronova}}, \bibinfo {author} {\bibfnamefont {M.~G.}\ \bibnamefont
  {Kozlov}},\ and\ \bibinfo {author} {\bibfnamefont {C.~W.}\ \bibnamefont
  {Clark}},\ }\href {https://doi.org/10.1103/PhysRevA.87.012509} {\bibfield
  {journal} {\bibinfo  {journal} {Phys. Rev. A}\ }\textbf {\bibinfo {volume}
  {87}},\ \bibinfo {pages} {012509} (\bibinfo {year} {2013})}\BibitemShut
  {NoStop}%
\bibitem [{\citenamefont {Middelmann}\ \emph {et~al.}(2012)\citenamefont
  {Middelmann}, \citenamefont {Falke}, \citenamefont {Lisdat},\ and\
  \citenamefont {Sterr}}]{PhysRevLett.109.263004}%
  \BibitemOpen
  \bibfield  {author} {\bibinfo {author} {\bibfnamefont {T.}~\bibnamefont
  {Middelmann}}, \bibinfo {author} {\bibfnamefont {S.}~\bibnamefont {Falke}},
  \bibinfo {author} {\bibfnamefont {C.}~\bibnamefont {Lisdat}},\ and\ \bibinfo
  {author} {\bibfnamefont {U.}~\bibnamefont {Sterr}},\ }\href
  {https://doi.org/10.1103/PhysRevLett.109.263004} {\bibfield  {journal}
  {\bibinfo  {journal} {Phys. Rev. Lett.}\ }\textbf {\bibinfo {volume} {109}},\
  \bibinfo {pages} {263004} (\bibinfo {year} {2012})}\BibitemShut {NoStop}%
\bibitem [{\citenamefont {Fasano}\ \emph {et~al.}(2021)\citenamefont {Fasano},
  \citenamefont {Chen}, \citenamefont {McGrew}, \citenamefont {Brand},
  \citenamefont {Fox},\ and\ \citenamefont {Ludlow}}]{2103.12052}%
  \BibitemOpen
  \bibfield  {author} {\bibinfo {author} {\bibfnamefont {R.}~\bibnamefont
  {Fasano}}, \bibinfo {author} {\bibfnamefont {Y.}~\bibnamefont {Chen}},
  \bibinfo {author} {\bibfnamefont {W.}~\bibnamefont {McGrew}}, \bibinfo
  {author} {\bibfnamefont {W.}~\bibnamefont {Brand}}, \bibinfo {author}
  {\bibfnamefont {R.}~\bibnamefont {Fox}},\ and\ \bibinfo {author}
  {\bibfnamefont {A.}~\bibnamefont {Ludlow}},\ }\href
  {https://doi.org/10.1103/PhysRevApplied.15.044016} {\bibfield  {journal}
  {\bibinfo  {journal} {Phys. Rev. Applied}\ }\textbf {\bibinfo {volume}
  {15}},\ \bibinfo {pages} {044016} (\bibinfo {year} {2021})}\BibitemShut
  {NoStop}%
\bibitem [{\citenamefont {Chou}\ \emph {et~al.}(2010)\citenamefont {Chou},
  \citenamefont {Hume}, \citenamefont {Rosenband},\ and\ \citenamefont
  {Wineland}}]{Science.329.1630}%
  \BibitemOpen
  \bibfield  {author} {\bibinfo {author} {\bibfnamefont {C.~W.}\ \bibnamefont
  {Chou}}, \bibinfo {author} {\bibfnamefont {D.~B.}\ \bibnamefont {Hume}},
  \bibinfo {author} {\bibfnamefont {T.}~\bibnamefont {Rosenband}},\ and\
  \bibinfo {author} {\bibfnamefont {D.~J.}\ \bibnamefont {Wineland}},\ }\href
  {https://doi.org/10.1126/science.1192720} {\bibfield  {journal} {\bibinfo
  {journal} {Science}\ }\textbf {\bibinfo {volume} {329}},\ \bibinfo {pages}
  {1630} (\bibinfo {year} {2010})}\BibitemShut {NoStop}%
\bibitem [{\citenamefont {Takano}\ \emph {et~al.}(2016)\citenamefont {Takano},
  \citenamefont {Takamoto}, \citenamefont {Ushijima}, \citenamefont {Ohmae},
  \citenamefont {Akatsuka}, \citenamefont {Yamaguchi}, \citenamefont
  {Kuroishi}, \citenamefont {Munekane}, \citenamefont {Miyahara},\ and\
  \citenamefont {Katori}}]{Nat.Photon.10.662}%
  \BibitemOpen
  \bibfield  {author} {\bibinfo {author} {\bibfnamefont {T.}~\bibnamefont
  {Takano}}, \bibinfo {author} {\bibfnamefont {M.}~\bibnamefont {Takamoto}},
  \bibinfo {author} {\bibfnamefont {I.}~\bibnamefont {Ushijima}}, \bibinfo
  {author} {\bibfnamefont {N.}~\bibnamefont {Ohmae}}, \bibinfo {author}
  {\bibfnamefont {T.}~\bibnamefont {Akatsuka}}, \bibinfo {author}
  {\bibfnamefont {A.}~\bibnamefont {Yamaguchi}}, \bibinfo {author}
  {\bibfnamefont {Y.}~\bibnamefont {Kuroishi}}, \bibinfo {author}
  {\bibfnamefont {H.}~\bibnamefont {Munekane}}, \bibinfo {author}
  {\bibfnamefont {B.}~\bibnamefont {Miyahara}},\ and\ \bibinfo {author}
  {\bibfnamefont {H.}~\bibnamefont {Katori}},\ }\href
  {https://doi.org/https://doi.org/10.1038/nphoton.2016.159} {\bibfield
  {journal} {\bibinfo  {journal} {Nat. Photon.}\ }\textbf {\bibinfo {volume}
  {10}},\ \bibinfo {pages} {662} (\bibinfo {year} {2016})}\BibitemShut
  {NoStop}%
\bibitem [{\citenamefont {Takamoto}\ \emph {et~al.}(2020)\citenamefont
  {Takamoto}, \citenamefont {Ushijima}, \citenamefont {Ohmae}, \citenamefont
  {Yahagi}, \citenamefont {Kokado}, \citenamefont {Shinkai},\ and\
  \citenamefont {Katori}}]{Nat.Photon.14.411}%
  \BibitemOpen
  \bibfield  {author} {\bibinfo {author} {\bibfnamefont {M.}~\bibnamefont
  {Takamoto}}, \bibinfo {author} {\bibfnamefont {I.}~\bibnamefont {Ushijima}},
  \bibinfo {author} {\bibfnamefont {N.}~\bibnamefont {Ohmae}}, \bibinfo
  {author} {\bibfnamefont {T.}~\bibnamefont {Yahagi}}, \bibinfo {author}
  {\bibfnamefont {K.}~\bibnamefont {Kokado}}, \bibinfo {author} {\bibfnamefont
  {H.}~\bibnamefont {Shinkai}},\ and\ \bibinfo {author} {\bibfnamefont
  {H.}~\bibnamefont {Katori}},\ }\href
  {https://doi.org/https://doi.org/10.1038/s41566-020-0619-8} {\bibfield
  {journal} {\bibinfo  {journal} {Nat. Photon.}\ }\textbf {\bibinfo {volume}
  {14}},\ \bibinfo {pages} {411} (\bibinfo {year} {2020})}\BibitemShut
  {NoStop}%
\bibitem [{\citenamefont {Itano}\ \emph {et~al.}(1993)\citenamefont {Itano},
  \citenamefont {Bergquist}, \citenamefont {Bollinger}, \citenamefont
  {Gilligan}, \citenamefont {Heinzen}, \citenamefont {Moore}, \citenamefont
  {Raizen},\ and\ \citenamefont {Wineland}}]{PhysRevA.47.3554}%
  \BibitemOpen
  \bibfield  {author} {\bibinfo {author} {\bibfnamefont {W.~M.}\ \bibnamefont
  {Itano}}, \bibinfo {author} {\bibfnamefont {J.~C.}\ \bibnamefont
  {Bergquist}}, \bibinfo {author} {\bibfnamefont {J.~J.}\ \bibnamefont
  {Bollinger}}, \bibinfo {author} {\bibfnamefont {J.~M.}\ \bibnamefont
  {Gilligan}}, \bibinfo {author} {\bibfnamefont {D.~J.}\ \bibnamefont
  {Heinzen}}, \bibinfo {author} {\bibfnamefont {F.~L.}\ \bibnamefont {Moore}},
  \bibinfo {author} {\bibfnamefont {M.~G.}\ \bibnamefont {Raizen}},\ and\
  \bibinfo {author} {\bibfnamefont {D.~J.}\ \bibnamefont {Wineland}},\ }\href
  {https://doi.org/10.1103/PhysRevA.47.3554} {\bibfield  {journal} {\bibinfo
  {journal} {Phys. Rev. A}\ }\textbf {\bibinfo {volume} {47}},\ \bibinfo
  {pages} {3554} (\bibinfo {year} {1993})}\BibitemShut {NoStop}%
\bibitem [{\citenamefont {Lemonde}\ and\ \citenamefont
  {Wolf}(2005)}]{PhysRevA.72.033409}%
  \BibitemOpen
  \bibfield  {author} {\bibinfo {author} {\bibfnamefont {P.}~\bibnamefont
  {Lemonde}}\ and\ \bibinfo {author} {\bibfnamefont {P.}~\bibnamefont {Wolf}},\
  }\href {https://doi.org/10.1103/PhysRevA.72.033409} {\bibfield  {journal}
  {\bibinfo  {journal} {Phys. Rev. A}\ }\textbf {\bibinfo {volume} {72}},\
  \bibinfo {pages} {033409} (\bibinfo {year} {2005})}\BibitemShut {NoStop}%
\bibitem [{\citenamefont {Sinha}\ and\ \citenamefont
  {Samuel}(2014)}]{ClassQuantumGrav.32.015018}%
  \BibitemOpen
  \bibfield  {author} {\bibinfo {author} {\bibfnamefont {S.}~\bibnamefont
  {Sinha}}\ and\ \bibinfo {author} {\bibfnamefont {J.}~\bibnamefont {Samuel}},\
  }\href {https://doi.org/10.1088/0264-9381/32/1/015018} {\bibfield  {journal}
  {\bibinfo  {journal} {Class. Quantum Grav.}\ }\textbf {\bibinfo {volume}
  {32}},\ \bibinfo {pages} {015018} (\bibinfo {year} {2014})}\BibitemShut
  {NoStop}%
\bibitem [{\citenamefont {McConnell}\ \emph {et~al.}(2015)\citenamefont
  {McConnell}, \citenamefont {Zhang}, \citenamefont {Hu}, \citenamefont
  {\'Cuk},\ and\ \citenamefont {Vuleti\'c}}]{Nature.519.439}%
  \BibitemOpen
  \bibfield  {author} {\bibinfo {author} {\bibfnamefont {R.}~\bibnamefont
  {McConnell}}, \bibinfo {author} {\bibfnamefont {H.}~\bibnamefont {Zhang}},
  \bibinfo {author} {\bibfnamefont {J.}~\bibnamefont {Hu}}, \bibinfo {author}
  {\bibfnamefont {S.}~\bibnamefont {\'Cuk}},\ and\ \bibinfo {author}
  {\bibfnamefont {V.}~\bibnamefont {Vuleti\'c}},\ }\href
  {https://doi.org/https://doi.org/10.1038/nature14293} {\bibfield  {journal}
  {\bibinfo  {journal} {Nature}\ }\textbf {\bibinfo {volume} {519}},\ \bibinfo
  {pages} {439} (\bibinfo {year} {2015})}\BibitemShut {NoStop}%
\bibitem [{\citenamefont {Baker}\ \emph {et~al.}(2009)\citenamefont {Baker},
  \citenamefont {Gillen}, \citenamefont {Peng}, \citenamefont {F\"olling},\
  and\ \citenamefont {Greiner}}]{Nature.462.74}%
  \BibitemOpen
  \bibfield  {author} {\bibinfo {author} {\bibfnamefont {W.~S.}\ \bibnamefont
  {Baker}}, \bibinfo {author} {\bibfnamefont {J.~I.}\ \bibnamefont {Gillen}},
  \bibinfo {author} {\bibfnamefont {A.}~\bibnamefont {Peng}}, \bibinfo {author}
  {\bibfnamefont {S.}~\bibnamefont {F\"olling}},\ and\ \bibinfo {author}
  {\bibfnamefont {M.}~\bibnamefont {Greiner}},\ }\href
  {https://doi.org/https://doi.org/10.1038/nature08482} {\bibfield  {journal}
  {\bibinfo  {journal} {Nature}\ }\textbf {\bibinfo {volume} {462}},\ \bibinfo
  {pages} {74} (\bibinfo {year} {2009})}\BibitemShut {NoStop}%
\bibitem [{\citenamefont {Sherson}\ \emph {et~al.}(2010)\citenamefont
  {Sherson}, \citenamefont {Weitenberg}, \citenamefont {Endres}, \citenamefont
  {Cheneau}, \citenamefont {Bloch},\ and\ \citenamefont
  {Kuhr}}]{Nature.467.68}%
  \BibitemOpen
  \bibfield  {author} {\bibinfo {author} {\bibfnamefont {J.~F.}\ \bibnamefont
  {Sherson}}, \bibinfo {author} {\bibfnamefont {C.}~\bibnamefont {Weitenberg}},
  \bibinfo {author} {\bibfnamefont {M.}~\bibnamefont {Endres}}, \bibinfo
  {author} {\bibfnamefont {M.}~\bibnamefont {Cheneau}}, \bibinfo {author}
  {\bibfnamefont {I.}~\bibnamefont {Bloch}},\ and\ \bibinfo {author}
  {\bibfnamefont {S.}~\bibnamefont {Kuhr}},\ }\href
  {https://doi.org/https://doi.org/10.1038/nature09378} {\bibfield  {journal}
  {\bibinfo  {journal} {Nature}\ }\textbf {\bibinfo {volume} {467}},\ \bibinfo
  {pages} {68} (\bibinfo {year} {2010})}\BibitemShut {NoStop}%
\bibitem [{\citenamefont {Cidrim}\ \emph {et~al.}(2021)\citenamefont {Cidrim},
  \citenamefont {Pi\~neiro Orioli}, \citenamefont {Sanner}, \citenamefont
  {Hutson}, \citenamefont {Ye}, \citenamefont {Bachelard},\ and\ \citenamefont
  {Rey}}]{PhysRevLett.127.013401}%
  \BibitemOpen
  \bibfield  {author} {\bibinfo {author} {\bibfnamefont {A.}~\bibnamefont
  {Cidrim}}, \bibinfo {author} {\bibfnamefont {A.}~\bibnamefont {Pi\~neiro
  Orioli}}, \bibinfo {author} {\bibfnamefont {C.}~\bibnamefont {Sanner}},
  \bibinfo {author} {\bibfnamefont {R.~B.}\ \bibnamefont {Hutson}}, \bibinfo
  {author} {\bibfnamefont {J.}~\bibnamefont {Ye}}, \bibinfo {author}
  {\bibfnamefont {R.}~\bibnamefont {Bachelard}},\ and\ \bibinfo {author}
  {\bibfnamefont {A.~M.}\ \bibnamefont {Rey}},\ }\href
  {https://doi.org/10.1103/PhysRevLett.127.013401} {\bibfield  {journal}
  {\bibinfo  {journal} {Phys. Rev. Lett.}\ }\textbf {\bibinfo {volume} {127}},\
  \bibinfo {pages} {013401} (\bibinfo {year} {2021})}\BibitemShut {NoStop}%
\bibitem [{\citenamefont {Dzuba}\ \emph {et~al.}(2018)\citenamefont {Dzuba},
  \citenamefont {Flambaum},\ and\ \citenamefont
  {Schiller}}]{PhysRevA.98.022501}%
  \BibitemOpen
  \bibfield  {author} {\bibinfo {author} {\bibfnamefont {V.~A.}\ \bibnamefont
  {Dzuba}}, \bibinfo {author} {\bibfnamefont {V.~V.}\ \bibnamefont
  {Flambaum}},\ and\ \bibinfo {author} {\bibfnamefont {S.}~\bibnamefont
  {Schiller}},\ }\href {https://doi.org/10.1103/PhysRevA.98.022501} {\bibfield
  {journal} {\bibinfo  {journal} {Phys. Rev. A}\ }\textbf {\bibinfo {volume}
  {98}},\ \bibinfo {pages} {022501} (\bibinfo {year} {2018})}\BibitemShut
  {NoStop}%
\bibitem [{\citenamefont {Sherman}\ \emph {et~al.}(2012)\citenamefont
  {Sherman}, \citenamefont {Lemke}, \citenamefont {Hinkley}, \citenamefont
  {Pizzocaro}, \citenamefont {Fox}, \citenamefont {Ludlow},\ and\ \citenamefont
  {Oates}}]{PhysRevLett.108.153002}%
  \BibitemOpen
  \bibfield  {author} {\bibinfo {author} {\bibfnamefont {J.~A.}\ \bibnamefont
  {Sherman}}, \bibinfo {author} {\bibfnamefont {N.~D.}\ \bibnamefont {Lemke}},
  \bibinfo {author} {\bibfnamefont {N.}~\bibnamefont {Hinkley}}, \bibinfo
  {author} {\bibfnamefont {M.}~\bibnamefont {Pizzocaro}}, \bibinfo {author}
  {\bibfnamefont {R.~W.}\ \bibnamefont {Fox}}, \bibinfo {author} {\bibfnamefont
  {A.~D.}\ \bibnamefont {Ludlow}},\ and\ \bibinfo {author} {\bibfnamefont
  {C.~W.}\ \bibnamefont {Oates}},\ }\href
  {https://doi.org/10.1103/PhysRevLett.108.153002} {\bibfield  {journal}
  {\bibinfo  {journal} {Phys. Rev. Lett.}\ }\textbf {\bibinfo {volume} {108}},\
  \bibinfo {pages} {153002} (\bibinfo {year} {2012})}\BibitemShut {NoStop}%
\end{thebibliography}%

\end{document}